\documentclass[aps]{revtex4}

\usepackage{graphicx}
\usepackage{latexsym}
\usepackage{graphicx}
\usepackage{epsfig}
\usepackage[english]{babel}
\usepackage[latin1] {inputenc}
\usepackage{amsmath,amsfonts,amssymb}
\usepackage{bm}
\usepackage[sort&compress]{natbib}
\newcommand \be {\begin{equation}}
\newcommand \bea {\begin{eqnarray}}
\newcommand \ee {\end{equation}}
\newcommand \eea {\end{eqnarray}}
\newcommand \bed {\begin{displaymath}}
\newcommand \eed {\end{displaymath}}

\newcommand{\bit}{\begin{itemize}}
\newcommand{\eit}{\end{itemize}}

\newcommand{\bgar}{\begin{eqnarray}}
\newcommand{\enar}{\end{eqnarray}}

\begin{document}

\centerline{\bf How macroscopic laws describe complex dynamics:}

\centerline{\bf asymptomatic population and CoviD-19 spreading}

\author{D. Lanteri~$^{(a,b,c,**)}$, D.~Carco'$^{(d)}$ and P.~Castorina$^{(a,b,*)}$ }
\affiliation{
\mbox{$^{(a)}$ INFN, Sezione di Catania, I-95123 Catania, Italy} \\
\mbox{$^{(b)}$ Faculty of Mathematics and Physics, Charles University} \\
\mbox{V Hole\v{s}ovi\v{c}k\'ach 2, 18000 Prague 8, Czech Republic} \\
\mbox{$^{(c)}$ Dipartimento di Fisica e Astronomia, Universit\`a di Catania, Italy}\\
\mbox{$^{(d)}$ Istituto Oncologico del Mediterraneo, Viagrande, Italy}
}

\date{\today}
\begin{abstract}
Macroscopic growth laws describe in an effective way the underlying complex dynamics of the spreading of infections, as in the case of CoviD-19, where the counting of the cumulative number $N(t)$ of detected infected individuals is a generally accepted  variable to understand the epidemic phase. However $N(t)$ does not take into account the unknown  number of asymptomatic cases $A(t)$. 
The considered model of CoviD-19 spreading is based on a system of coupled differential equations, which include the dynamics of the spreading among symptomatic and asymptomatic individuals and the strong containment effects due to the social isolation. The solution has ben compared with $N(t)$, determined by a single differential equation with no explicit reference to $A(t)$, showing the equivalence of the two methods.  The model is applied to Covid-19 spreading in Italy where a transition from an exponential behavior to a Gompertz growth for $N(t)$ has been observed in more recent data.
The information contained in the time series  $N(t)$  turns out to be reliable to understand the epidemic phase, although it does not describe the total infected population. The asymptomatic population is larger than the symptomatic one in the fast growth phase of the spreading.

\end{abstract}
% \pacs{89.75.-k} 
 \maketitle

\section{Introduction}

There is an impressive number of experimental verifications, in many different scientific sectors, that coarse-grain properties of systems, with simple laws with respect to the fundamental microscopic algorithms, emerge at different levels of magnification providing important tools for explaining and predicting new phenomena.

For example, the Gompertz law (GL)~\cite{GL}, initially applied to human mortality tables (i.e. aging), describes tumor growth~\cite{steel,norton}, kinetics of enzymatic reactions, oxygenation of hemoglobin, intensity of photosynthesis as a function of CO2 concentration, drug dose-response curve, dynamics of growth, (e.g., bacteria, normal eukaryotic organisms). Analogously, the Logistic Law (LL)~\cite{LL}   has been applied in population dynamics, in economics, in material science and in many other sectors.

The ability of macroscopic growth laws in describing the underlying complex dynamics is sometime surprising. 

A clear and timely example is given by the infection spreading  (Coronavirus \cite{oms1,oms2} in particular), where different  Governments impose strong containment efforts on the basis of the mortality rate and of the  growth rate of the cumulative total number of detected infected people $N(t)$ at time $t$, which has a strong impact on the national health systems.

However, there is a large number of infected people without any symptoms  who contribute to the disease spreading but are not explicitly taken into account in $N(t)$, since not detected.

More precisely, a macroscopic growth law for $N(t)$ is solution of the general differential equation
\be\label{eq:1}
\frac{dN}{dt} = \alpha_{eff}(t) N(t),
\ee
where $\alpha_{eff}(t)$ is the specific growth rate at time $t$. For example, if $\alpha_{eff(t)}=$constant one obtains the exponential behavior.  Various growth patterns  have been very recently applied to the time evolution of the CoviD-19 infection \cite{np7,np8,np9,np10,np11,np12,np13}. On the other hand,  by using the previous equation to describe the epidemic phase,  the cumulative number of asymptomatic infected individuals, $A(t)$, seems completely neglected.  Indeed, the second term of the equation depends on $N(t)$ only.

For the  CoviD-19 infection, since $A(t)$ is unknown,  many Governments have correctly applied strong constraints to slow down the spreading: social isolation, information on the localization of infected individuals and the use of a very large number of medical swabs.

However, the question arises: is the information obtained by monitoring $N(t)$ reliable in understanding  the epidemic phase?

By a simple model of the interaction between the symptomatic cumulative detected population $N(t)$ and the asymptomatic one, $A(t)$, we discuss how the effective coarse-grain equation (1)  takes into account the asymptomatic population in the transient phase of the spreading. Moreover one gets useful indications on the number of asymptomatic individuals and on the effective lethality.

In Sec.1  different macroscopic growth laws are applied to the cumulative number of detected infected individuals in China, South Korea and Italy, where the containment effort
started earlier, to describe the phases of the spreading. A simple dynamical model where the asymptomatic population and the isolation effect play an important role is discussed in Sec.2.
The results are in Sec.3 and Sec.4 is devoted to comments and conclusions.

\section{Infection spreading phases: macroscopic description}

The macroscopic growth laws in eq.~\eqref{eq:1}, can be classified by considering the time derivative of $\alpha_{eff}$, as shown in ref.~\cite{cast1,cast2}. For example the GL and the LL are in the, so called, universality class U2~\cite{cast1}, where $d\alpha_{eff}/dt= a_1 \alpha+a_2 \alpha^2$. Moreover, the Gompertz  and the logistic equations can be written in a different  way, which clarifies the feedback effect of the increasing population , i.e.
\be\label{eq:G}
\frac{1}{N(t)}\frac{dN(t)}{dt} = - k_g\;\ln \frac{N(t)}{N_\infty^g}  \qquad \text{Gompertz}\;,
\ee
\be
\frac{1}{N(t)}\frac{dN(t)}{dt} =  k_l \left(1- \frac{N(t)}{N_{\infty}^l}\right)
\qquad 
\text{logistic}
\ee
where $k_g$, $N_\infty^g$, $k_l$ and $N_\infty^l$ are constants.

In the GL and in the LL, $k_g \ln(N_\infty^g)$ and $k_l$ are respectively the initial exponential rates and the other terms determine their slowdown.

\begin{figure}
	\centering
	\includegraphics[width=0.8\columnwidth]{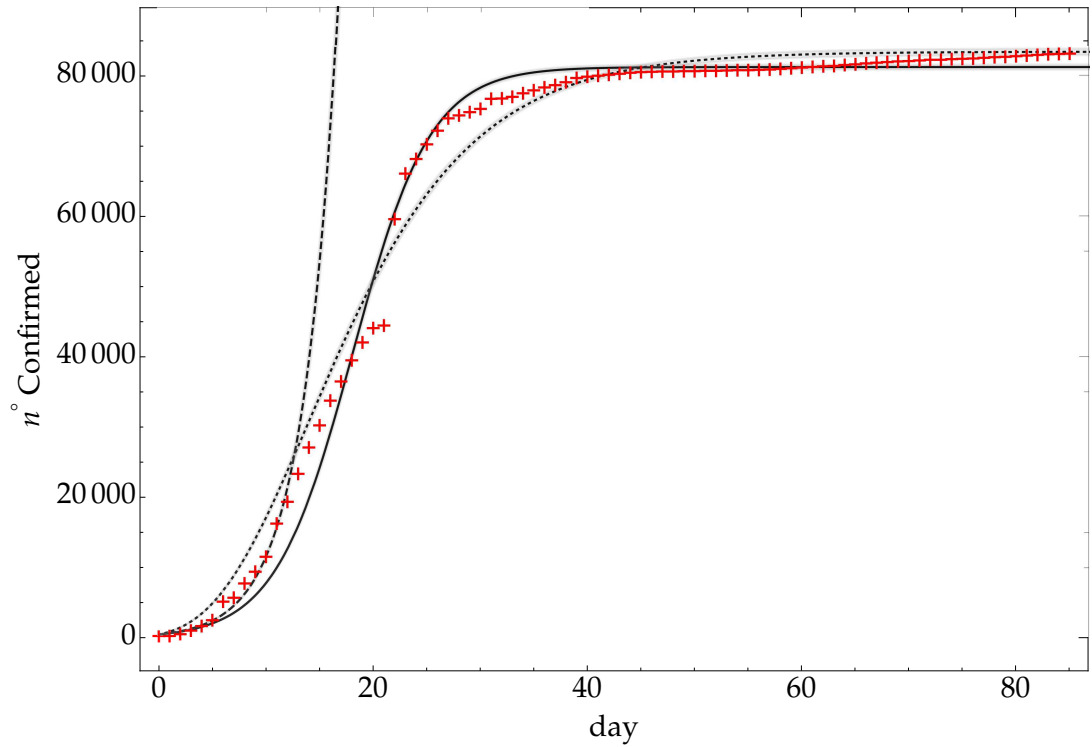}
	\caption{Comparison of the growth laws with the data of the cumulative number of infected individuals in China: continuous line si the logistic curve, dotted line the Gompertz, the dashed one the exponential. Time zero corresponds to the initial day - 22/01.}
	\label{fig:China}
\end{figure}

\begin{figure}
	\centering
	\includegraphics[width=0.8\columnwidth]{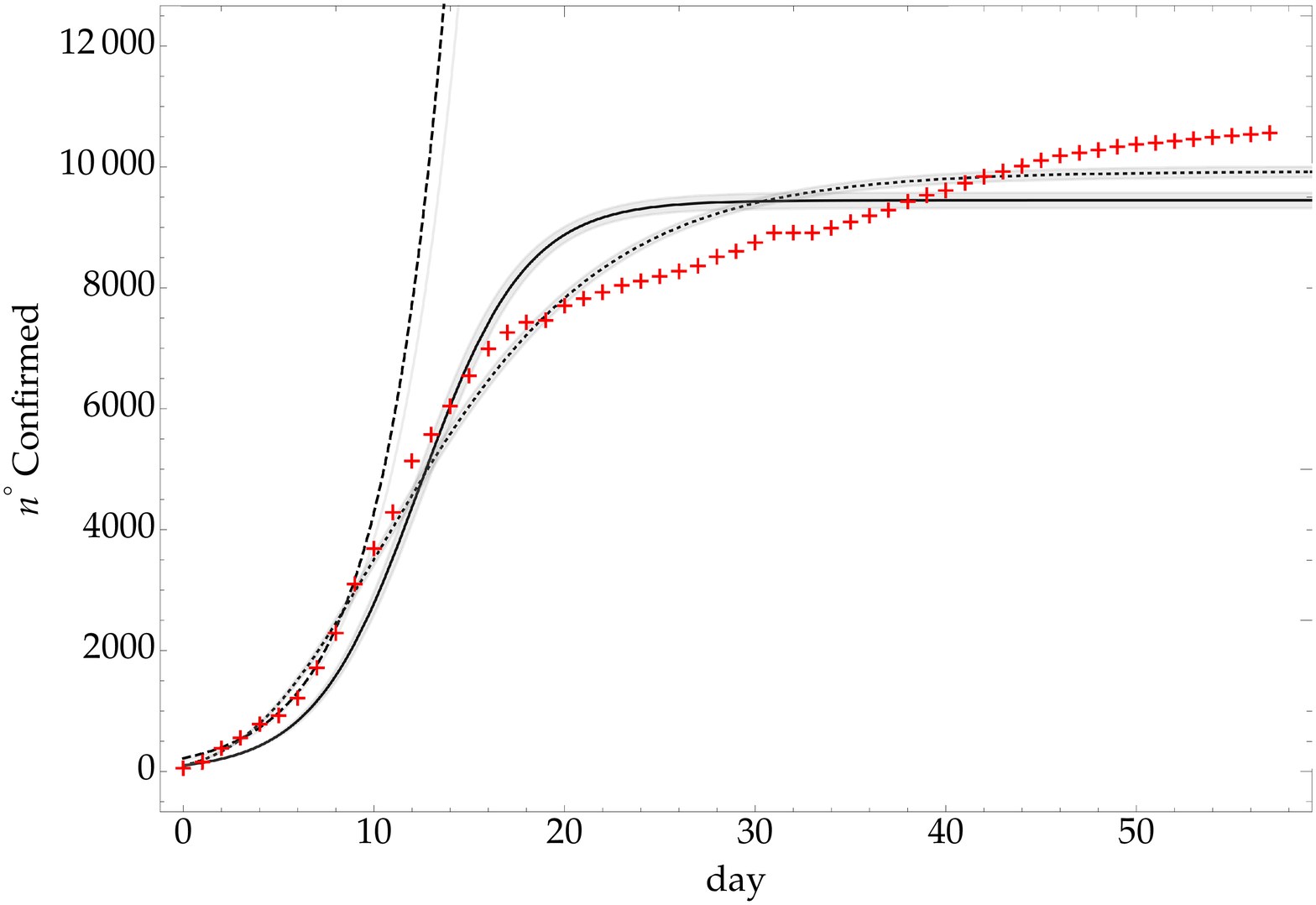}
	\caption{Comparison of the growth laws with the data of the cumulative number of infected individuals in South Korea: continuous line si the logistic curve, dotted line the Gompertz, the dashed one the exponential. Time zero corresponds to the initial day - 20/02.}
	\label{fig:SK}
\end{figure}

\begin{figure}
	\centering
	\includegraphics[width=0.8\columnwidth]{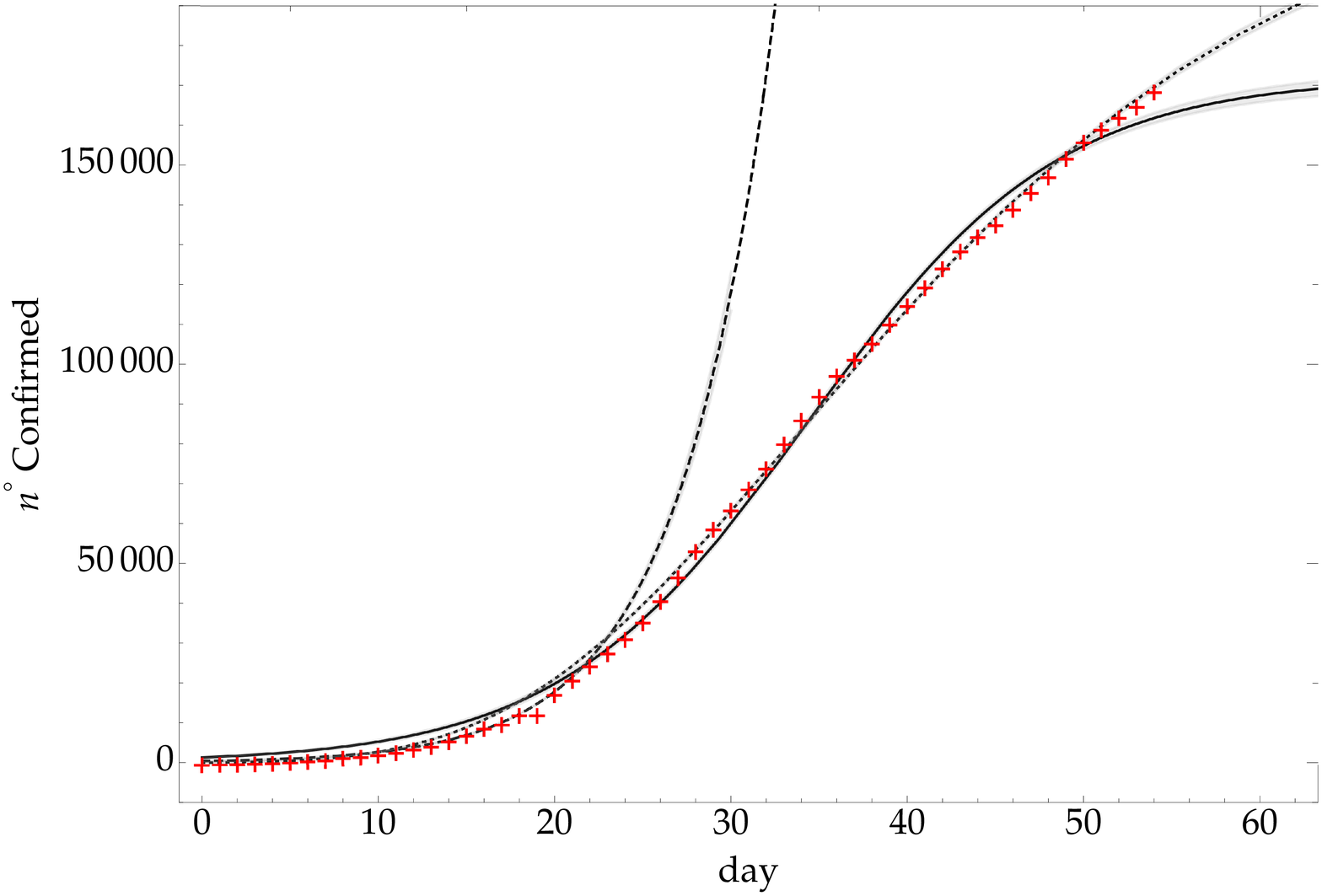}
	\caption{Comparison of the growth laws with the data of the cumulative number of infected individuals in Italy: continuous line si the logistic curve, dotted line the Gompertz, the dashed one the exponential. Time zero corresponds to the initial day - 22/02.}
	\label{fig:Italy}
\end{figure}

The comparison of the growth laws, solutions of the previous differential equations, with the data on the cumulative number of infected individuals is reported in fig.~\ref{fig:China} for China,
showing that the coronavirus spreading has three phases: an initial exponential behavior, followed by a Gompertz one and a final logistic phase. 
For South Korea, after the three phases, the spreading seems to restart (see fig.~\ref{fig:SK}). Italy is still in the Gompertz growth phase (see 
fig. \ref{fig:Italy}).

The previous equations  neglect the large (undetected) asymptomatic population and the isolation effects, since they concern the time evolution of $N(t)$ only.
However, a simple model in the next section shows that the GL takes into account a more complex underlying dynamics. The analysis can be repeated for the LL.

\section{A simple model: asymptomatic population and containment effects}

In the model, we call $T(t)$ the cumulative total number of infected people, which is the sum of the  number of the cumulative detected infected individuals $N(t)$ and of the asymptomatic ones $A(t)$: $T(t) = N(t) + A(t)$.

In the data of $N(t)$ the number of dead and of healed people is included, since they have been previously infected. In the fast growing phase, their total number is much smaller than $N(t)$ and therefore  they are not included in the dynamics (the re-infection is considered very unlikely). The previous condition defines the transient phase. 

Therefore one has to take into account the rates of the following processes ($n$ = detected infected person, $a$ = asymptomatic infected person): 
$ n \rightarrow n +n$, $a \rightarrow n +a$, $n \rightarrow n +a$, $a \rightarrow a + a$ with rates $c_1(t)$, $c_2(t)$, $c_3(t)$, $c_4(t)$  respectively.

Accordingly, the corresponding mean field equations are
\be\label{eq:A}
\frac{dN(t)}{dt} = c_1(t) N(t) + c_2(t) A(t),
\ee
\be\label{eq:B}
\frac{dA(t)}{dt} = c_3(t) N(t) + c_4(t) A(t). 
\ee
The functions  $c_i(t)$, $i = 1, 4$, describe the damping effects due to the containment effort and  we assume  they have an exponential decreasing behavior:
\be
c_i(t) \simeq e^{-\lambda t}
\ee
i.e. the rates of the dynamical  processes decrease in time due to social isolation and other external constraints. This effect should be independent on the status of the infected individual (n or a). On the other hand, in the  processes involving a symptomatic individual, he/she is (or should be) rapidly detected, together with those who belong to his/her chain of infection transmission, independently on their symptomatic or asymptomatic condition. Therefore the rate of the processes involving infected  persons  should be suppressed with respect to the transmission rate  among asymptomatic persons,  which is ``invisible''. 

In other terms, one expects that, in the transient phase, the rate of the process among asymptomatic  individuals only, $a \rightarrow a + a$, decreases slowly than the other ones. 
As a first step,  one assumes that the functions $c_1(t)$, 
$c_2(t)$, $c_3(t)$, $c_4(t)$ are given by
\be\label{eq:C}
c_1(t)=c_1^0 e^{-\lambda_1 t}, \phantom{...} c_2(t)= c_2^0 e^{-\lambda_1 t}, \phantom{...} c_3(t) = c_3^0 e^{-\lambda_1 t}, \phantom{...} c_4(t)=c_4^0 e^{-\lambda_2 t},  
\ee
with $\lambda_2 < \lambda_1$ and $c_i^0 > 0$. The coefficients $c_i^0$ are the initial rates and $\lambda_1$, $\lambda_2$ parameterize their reduction due to the isolation conditions.

 it is an effective way to take into account the dynamics in eqs.~(\ref{eq:A},~\ref{eq:B},~\ref{eq:C}).

To clarify that the solution of eq.(2) takes into account, in an effective way, the dynamics  in eqs.~(\ref{eq:A},~\ref{eq:B},~\ref{eq:C}), the system is analytically solved (see appendix)
for $\lambda_1=\lambda_2$.  For $\lambda_2 < \lambda_1$   there is no analytical solution and the  numerical one will be  compared with the Italian data and the previous GL fit in the next section.

\section{Results}

\begin{figure}
	\centering
	\includegraphics[width=0.8\columnwidth]{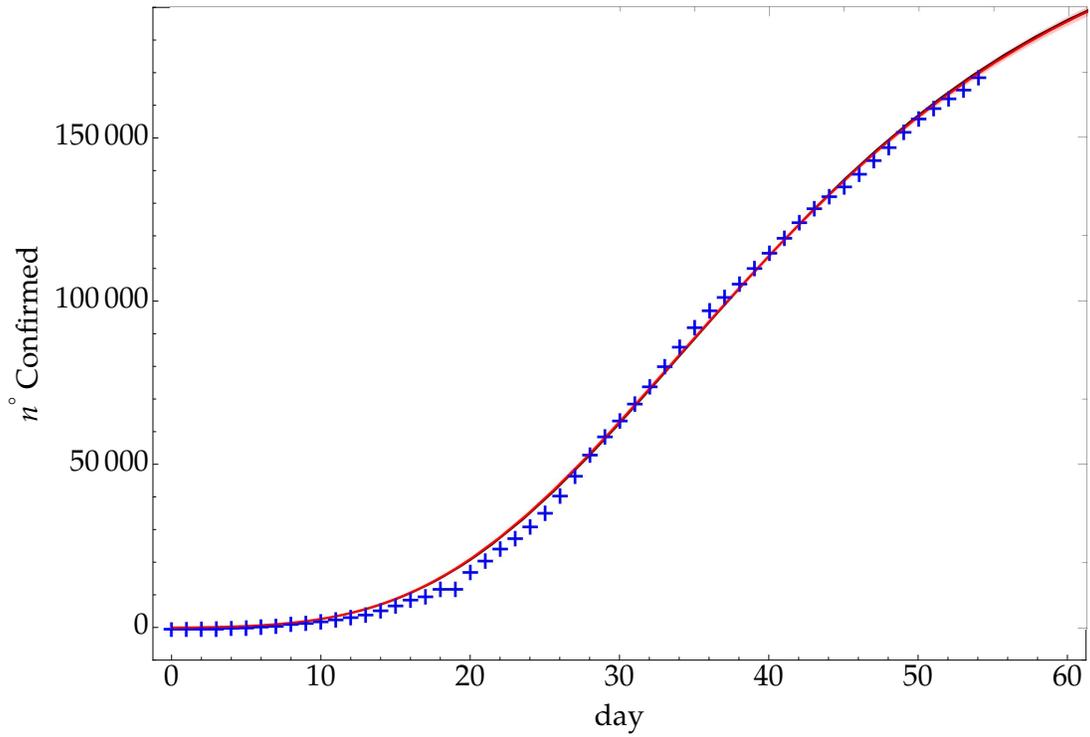}
	\caption{Cumulative  number of infected individuals in Italy from February  22nd (day 0) to April the 16th.  The GL fit (red) and the solution of the system in eq.~(\ref{eq:A},~\ref{eq:B},~\ref{eq:C}) (black) are plotted versus the observed data. The parameters are:  $N(0)=62$, $A(0)=N(0)$, $c_1^0=0.23$, $c_2^0=0.26$, $c_3^0=0.24$, $c_4^0=0.32$, $\lambda_1=0.0732$, $\lambda_2=0.045$. The two curves overlap.}
	\label{fig:B}
\end{figure}

\begin{figure}
	\centering
	\includegraphics[width=0.8\columnwidth]{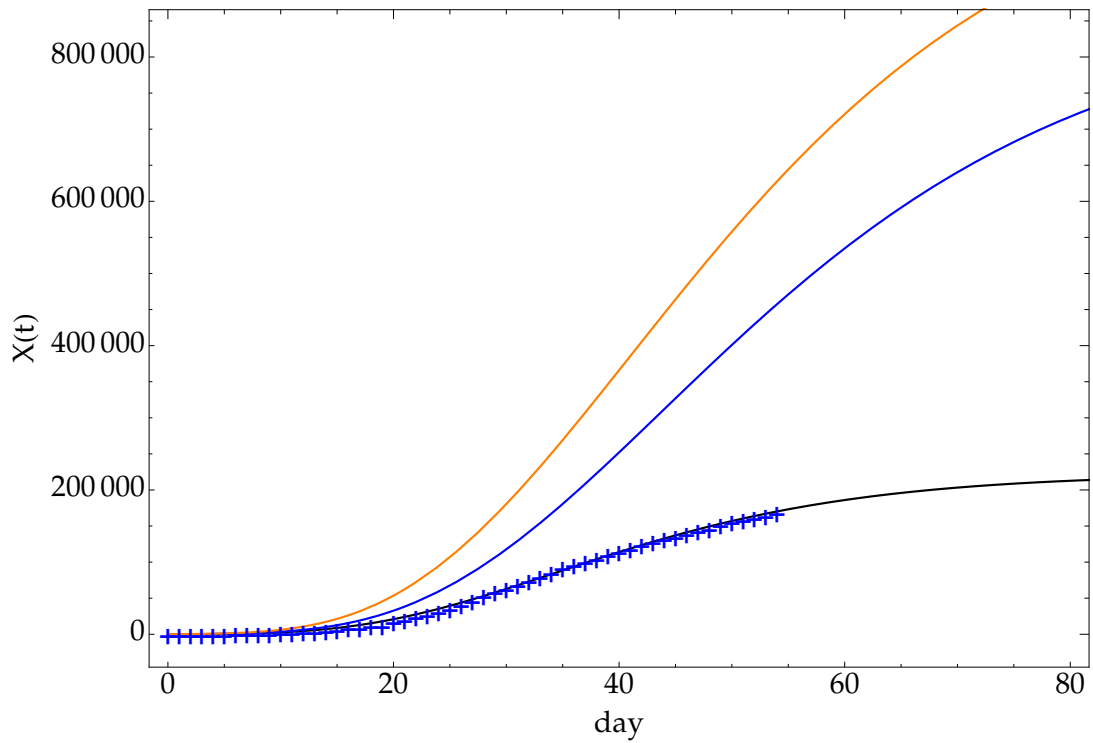}
	\caption{Cumulative number of infected individuals in Italy from February 22nd(day 0) to March 22nd.  The  solution of the system in eqs.~(\ref{eq:A},~\ref{eq:B},~\ref{eq:C})  are plotted: detected in black, asymptomatic in blue,  total in orange. }
		\label{fig:C}
\end{figure}

The Gompertz fit (from eq.~\eqref{eq:G}) of Italian data is  compared with the numerical solution of the dynamical system~(\ref{eq:A},~\ref{eq:B},~\ref{eq:C}) with $\lambda_2 < \lambda_1$ in fig.~\ref{fig:B}. The two curves are well compatible, indicating that the interaction between symptomatic and asymptomatic populations is  contained in eq.~\eqref{eq:G} in an effective way. Moreover the solution of the system gives an estimate of the asymptomatic population, plotted in fig.~\ref{fig:C}. 

The previous analysis clearly suggests that:
\begin{itemize}
	\item[a)] the curve $N(t)$ takes into account the dynamics of the asymptomatic individuals;
	
	\item[b)] in the fast growing phase there is a large asymptomatic population: to stop the infection spreading the strong social isolation is  the best method;
	
	\item[c)] the time evolution of $N(t)$ is an effective  macroscopic growth laws solution of eq.~\ref{eq:G} for Italy, completely consistent with the system in eqs.~(\ref{eq:A},~\ref{eq:B},~\ref{eq:C}).
\end{itemize}
In this respect the choice of monitoring the time evolution of $N(t)$ to understand the epidemic phase is reliable, since the
underling dynamics is included in the time dependence of $\alpha_{eff}$ in eq.~\eqref{eq:1}. This is probably related to the result 
that a general classification of the growth laws  depends on the time derivative of $\alpha_{eff}$~\cite{cast1,cast2}.

\begin{figure}
	\centering
	\includegraphics[width=0.8\columnwidth]{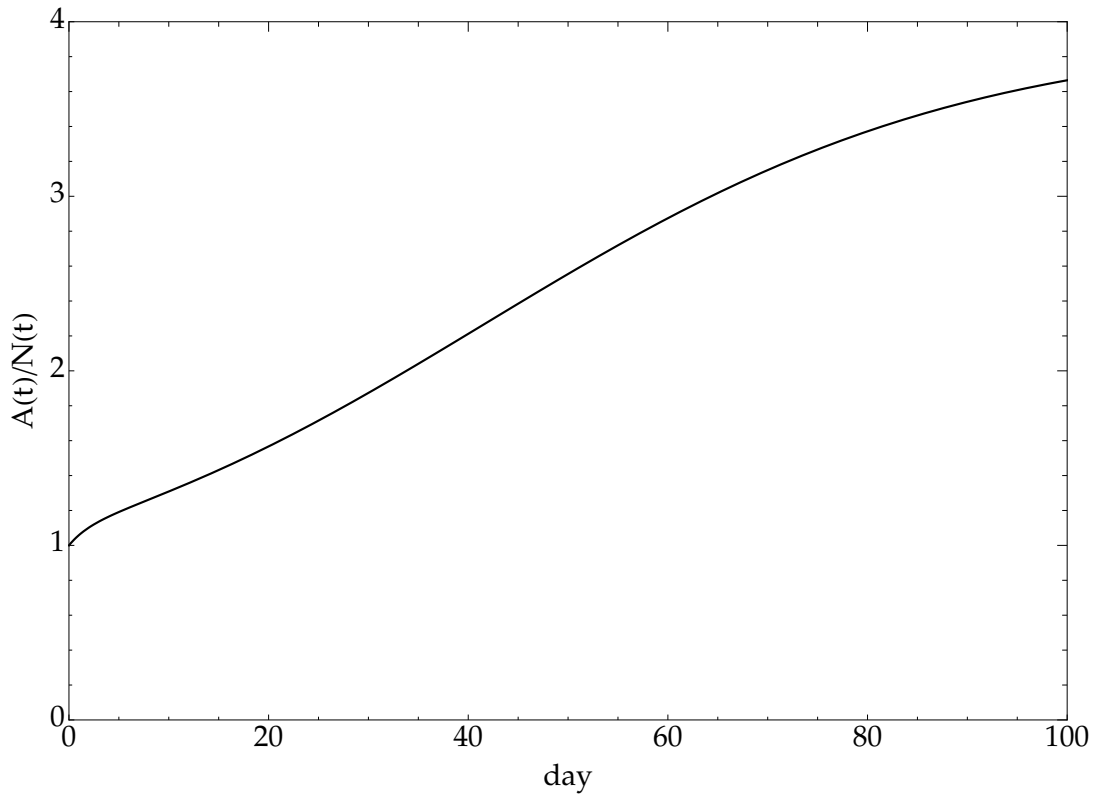}
	\caption{Ratio between asymptomatic and symptomatic populations.}
	\label{fig:R}
\end{figure}

On the other hand, the system of differential equations give more information than the single equation for $N(t)$. In particular, the ratio, $R(t)$ (see appendix),  between symptomatic and asymptomatic populations is plotted as a function of time in fig.~\ref{fig:R}. Moreover, one can evaluate in a more reliable way the lethality of the infection, defined as the ratio of the number of  dead individuals and the total number of infected ones. The italian lethality is $13\%$ neglecting the asymptomatic population, but, in our model, its value reduces to  $\simeq 3 \%$ considering the total infected population.

\section{Comments and Conclusions}

The model in eqs.~(\ref{eq:A},~\ref{eq:B},~\ref{eq:C}) is a simplified version of a more complex dynamics~\cite{net1,net2} and  can be improved including other specific populations~\cite{smaller}.

Different analyses of  the CoviD-19 spreading predict a large asymptomatic population.  In ref.~\cite{np10} the number of asymptomatic individuals in Italy on March 12th  turns out to be more than 100.000 versus a symptomatic detected population of about 12.000. Our analysis suggests an asymptotic population of about 27500 individuals on March  12th, 76000 on March  19th and 129000 on March 24th.  In ref.~\cite{lan}, the outbreak in Italy has been estimated, giving a ratio of about 1:3 between symptomatic and asymptomatic individuals on February 29Th. The evaluation of the asymptomatic population is a difficult task and the results are strongly model dependent.

Macroscopic laws in eq.~\eqref{eq:1}  include the dynamics of the infection  with the advantage of a reduction of the free parameters: the simple system in eqs.~(\ref{eq:A},~\ref{eq:B},~\ref{eq:C}) has in principle 5 parameters plus 2 initial conditions, but the  Gompertz curve has 2 parameters and 1 initial condition. 

The reliability of $N(t)$ as an index of the spreading has  been checked by a devoted analysis, including  the crucial  effects of the  social isolation and the role of the asymptomatic individuals.

\begin{figure}
	\centering
	\includegraphics[width=0.8\columnwidth]{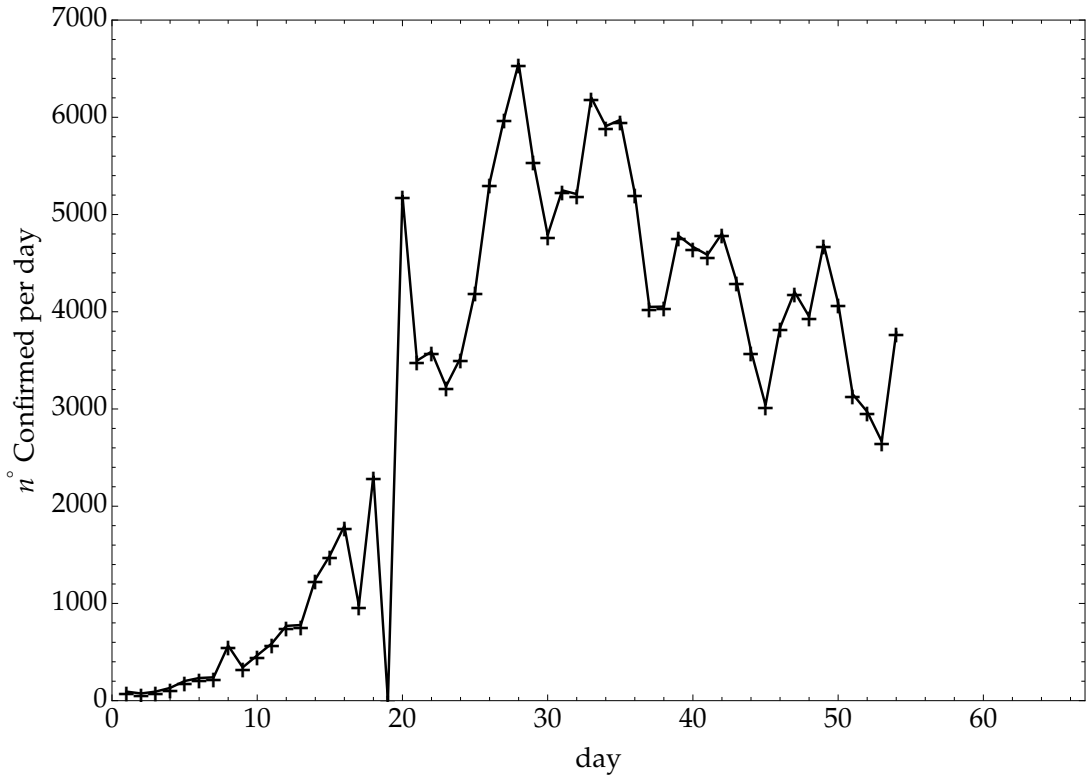}
	\caption{Confirmed daily rate.}
	\label{fig:ConfDay}
\end{figure}

\begin{figure}
	\centering
	\includegraphics[width=0.8\columnwidth]{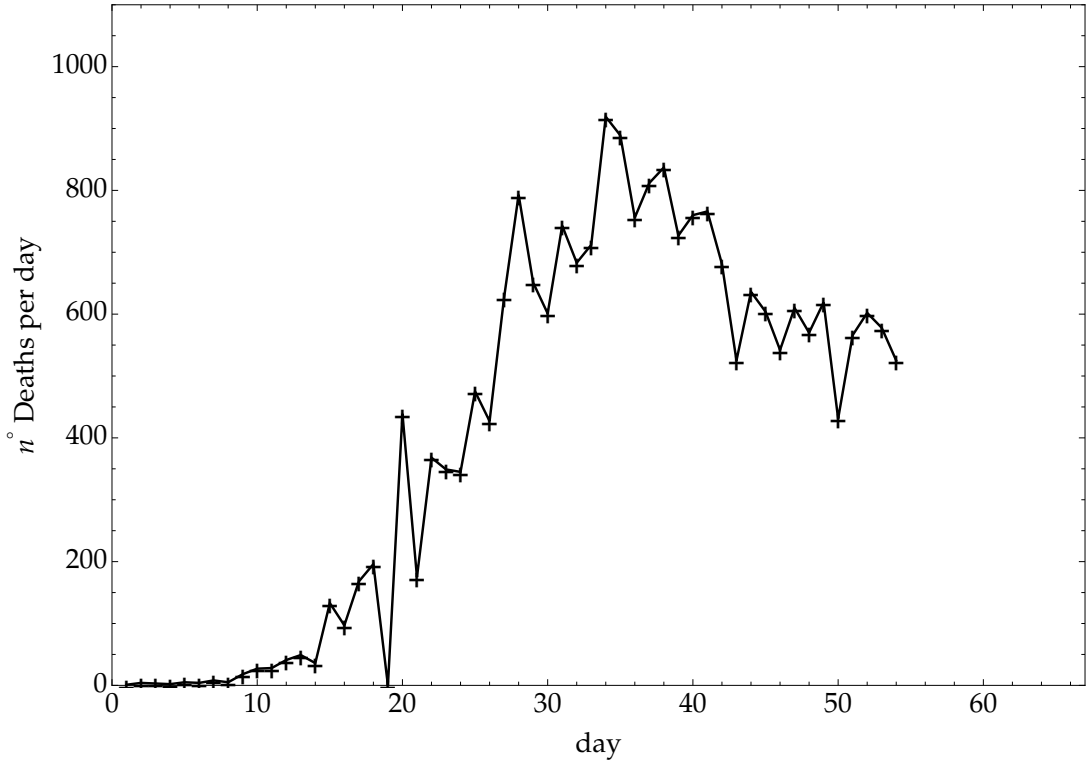}
	\caption{Mortality daily rate.}
	\label{fig:DDay}
\end{figure}

A further confirmation of this conclusion comes from the direct comparison between the  daily rate $N(t+1)-N(t)$ and the corresponding mortality one, reported in figs.~(\ref{fig:ConfDay},\ref{fig:DDay}), which shows that the time evolution of $N(t)$ anticipates the other by about 8 days.

It should be clear that the previous analysis implicitly assumes that the protocol concerning the detection of infected individuals, as for example  the number of medical swabs (per day),  does not change in time. Indeed, a modification in the detecting protocol (as in the Chinese case) or the sudden, strong, variation in the number of swabs could mimic an increase or a decrease in $N(t)$ of not dynamical origin.

\section{Acknowledgements}
P.C. thanks V. Latora for useful comments and suggestions.

\section{Appendix}

Let us consider the system of differential equations 
\begin{equation}\label{eq:s}
\begin{cases}
\displaystyle \frac{dN(t)}{dt} = c_1(t) N(t) + c_2(t) A(t)
\\\\
\displaystyle \frac{dA(t)}{dt} = c_3(t) N(t) + c_4(t) A(t)
\end{cases}
\;,
\end{equation}
with 
\begin{equation}\label{1}
c_i = c_i^0 \; e^{-\lambda \; t}\;,
\end{equation}.

It can be  solved by defining the ratio 
\begin{equation}
R(t) \equiv \frac{A(t)}{N(t)} 
\;,
\end{equation}
which satisfies the  equation 
\begin{equation}\label{eq:R}
\frac{dR(t)}{dt}
=
e^{-\lambda\;t} \left[c_3^0-R(t)\;\left[c_1^0-c_4^0+c_2^0\;R(t)\right]\right]
\;.
\end{equation}
The analytical solution of the previous eq.~\eqref{eq:R} is
\begin{equation}
R(t) 
=
\frac{R_+ - R_-\;\frac{R_0-R_+}{R_0-R_-}\;e^{-\frac{c_2^0}{\lambda}\;(R_+ - R_-)\;\left(1-e^{-\lambda \; t}\right)}}{1-\frac{R_0-R_+}{R_0-R_-}\;e^{-\frac{c_2^0}{\lambda}\;(R_+ - R_-)\;\left(1-e^{-\lambda \; t}\right)}}  
\;,
\end{equation}
where
\begin{equation}
R_\pm 
= 
\frac{1}{2}\left[-\frac{c_1^0  -c_4^0}{c_2^0} \pm \sqrt{\left(\frac{c_1^0  -c_4^0}{c_2^0}\right)^2 + \frac{4\;c_3^0}{c_2^0} }\;\right]
\end{equation}
and $R_0=R(t=0)=A_0/N_0$, being $N_0$ and $A_0$ the number of detected and asymptomatic infected individuals at the initial time respectively.

\begin{figure}
	\centering
	\includegraphics[width=0.8\columnwidth]{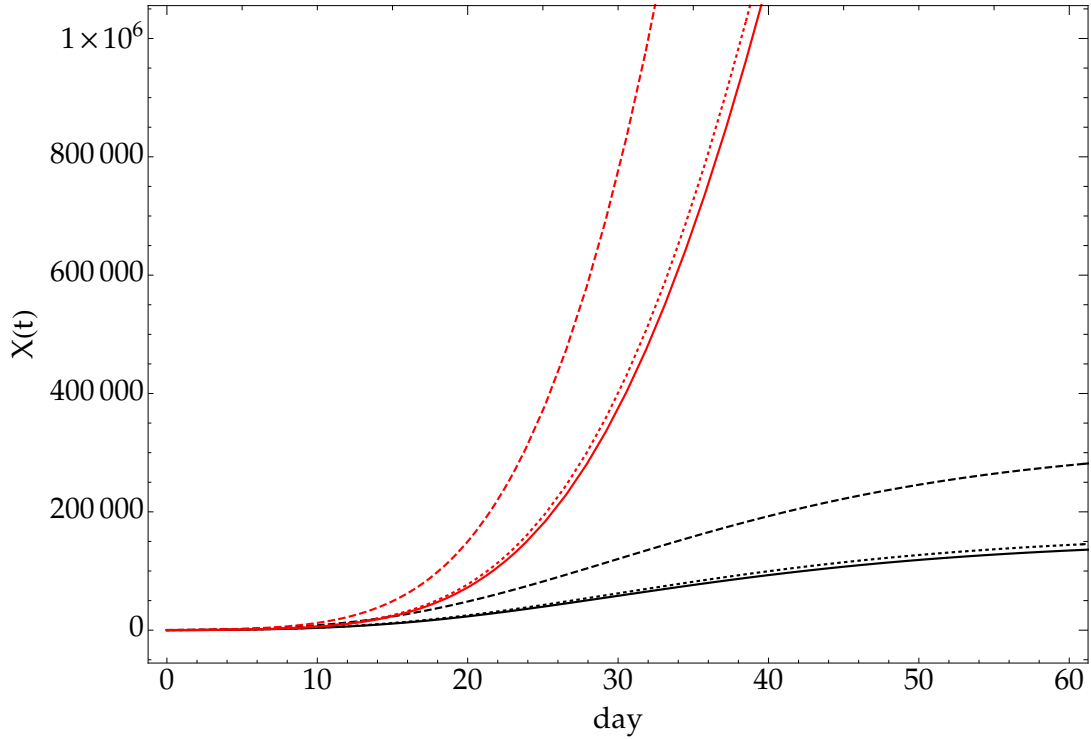}
	\caption{Solution of eq.~\eqref{eq:s} with $N(0)=62$, $A(0)=N(0)$, $c_1^0=0.23$, $c_2^0=0.26$, $c_3^0=0.24$, $c_4^0=0.32$. Black curves are for $\lambda=0.0732$, the red ones for  $\lambda=0.045$. Continuous curves give the cumulative number of detected infected
		individuals ($N(t)$), the dotted one correspond to the asymptomatic ones, $A(t)$, and the dashed  line is the total.}
	\label{fig:App}
\end{figure}

Putting $A(t) = R(t)\;N(t)$ in eq.~\eqref{eq:s} it turns out that 
\begin{equation}\label{eq:N}
N(t)
=
\frac{N_0\;e^{\frac{c_1^0+c_2^0\;R_+}{\lambda}\left(1-e^{-\lambda t}\right)}}{R_--R_+}\;
\left[(R_0-R_+)\;e^{\frac{c_2^0}{\lambda} (R_--R_+) \left(1-e^{-\lambda t}\right)}-R_0+R_-\right]
\;,
\end{equation}
and 
\begin{equation}
A(t)
=
\frac{N_0\;e^{\frac{ c_1^0+c_2^0\;R_+}{\lambda}\left(1-e^{-\lambda t}\right)}}{R_+-R_-}
\Bigg[
R_-\;(R_+-R_0)\;e^{\frac{c_2^0 (R_--R_+)}{\lambda}  \left(1 -e^{-\lambda t}\right)}
 +
R_+ (R_0-R_-)\Bigg]
\;.
\end{equation}
Finally, for the specific case  
$c_3^0=c_1^0 $ and $c_2^0=c_4^0 = k\;c_1^0$, with $k$ some constant,
\begin{equation}\label{eq:a}
N(t)
=
\frac{N_0}{k+1}\;\left((k\;R_0+1) e^{\frac{c_1^0(k+1)}{\lambda} \;\left(1-e^{-\lambda t}\right)}+k (1-R_0)\right)
\;,
\end{equation}
\begin{equation}
A(t)
=
\frac{N_0}{k+1}
\; \left((k\;R_0+1)\;e^{\frac{c_1^0 (k+1)}{\lambda}\;\left(1-e^{-\lambda t}\right)}+R_0-1\right)
\;,
\end{equation}
and
\begin{equation}
R(t)
=
1+
\frac{(R_0-1)(1+k)}{(k\;R_0+1)\;e^{\frac{c_1^0\; (k+1)}{\lambda}\; \left(1-e^{-\lambda  t}\right)}+k (1-R_0)}
\end{equation}
The typical time dependence of the previous solutions, $N(t)$ and $A(t)$, is plotted in figure~\ref{fig:App}.

\end{document}